\documentstyle[prl,aps,floats,twocolumn]{revtex}
\begin{document}
\def\be{\begin{equation}}
\def\ee{\end{equation}}
\def\bea{\begin{eqnarray}}
\def\eea{\end{eqnarray}}
\def\E{{\rm e}}
\def\bearst{\begin{eqnarray*}}
\def\eearst{\end{eqnarray*}}
\def\peleven{\parbox{11cm}}
\def\peffec{\peight{\bearst\eearst}\hfill\peleven}
\def\pspace{\peight{\bearst\eearst}\hfill}
\def\ptwelve{\parbox{12cm}}
\def\peight{\parbox{8mm}}
\twocolumn[\hsize\textwidth\columnwidth\hsize\csname@twocolumnfalse\endcsname 

\title
{ Multi-scale Correlation Functions in Strong Turbulence}
\author
{ Jahanshah Davoudi $^{a,c,d}$ and M. Reza Rahimi Tabar$^{b,c,d}$}
\address
{\it $^a$ Dept. of Physics , Sharif University of Technology,
P.O.Box 11365-9161, Tehran, Iran.\\
$^b$ CNRS UMR 6529, Observatoire de la C$\hat o$te d'Azur,
BP 4229, 06304 Nice Cedex 4, France,\\
$^c$ Dept. of Physics , Iran  University of Science and Technology,
Narmak, Tehran 16844, Iran.\\
$^d$ Institue for Studies in Theoretical Physics and 
Mathematics
Tehran P.O.Box: 19395-5746, Iran,\\
}  


\maketitle
\begin{abstract}

Under the framework of V. Yakhot {[Phys.Rev.E, {\bf57}, 1737 (1998)]} 
modelling of intermittent  structure functions in fully developed
turbulence and based on the experimentally supported Markovian nature 
of turbulence cascades 
{[R.Friedrich and J.Peinke, Phys.Rev.Lett, {\bf 78}, 863 (1997)]}, 
we calculate the multiscaling correlation functions. 
Fusion rules {[V.S.L'vov and I. Procaccia, Phys.Rev.Lett. {\bf 76}, 2898
(1996)]} which are experimentally tested 
{[R. Benzi, L. Biferale, and F. Toschi, Phys.Rev.Lett. {\bf 80}, 3244
(1998)]} to be compatible with almost uncorrelated multiplicative process
can analyticaly be supported by direct
calculations.

PACS numbers, 47.27.Ak, 47.27.Gs, 47.27.Jv and 47.40.Ki.

\end{abstract}


\hspace{.2in}
]

One of the most important issues in stationary turbulence is the intermittent
behaviour of the velocity fluctuations in the inertial range.
Understanding the statistical properties of intermittency is one of the most
challenging open problems in three dimensional fully developed turbulence.
The structures that arise in a random flow of stationary turbulence resembel
themselves as high peaks at random places and random times. The intervals 
between them are characterised by a low intensity and a large size. Rare peaks 
are the hallmarks of PDF's non-guassian tails.  
These strongly non-Gausssian activities are statistically scale-invariant
processes responsible for energy transfer.
Intermittency in the inertial range is usually analysed by means of the 
statistical properties of velocity differences,
$\delta_{r}u(x)=u(x+r)-u(x)$ \cite{fri}.
The overwhelming experimental and theoretical works have been brought forward 
for characterisation of structure functions; i.e. $S_{p}=\langle (\delta_{r}u(x)) ^{p} \rangle$.
A wide agreement exists on the fact that $S_{p}(r)$ exhibits a scaling behaviour in the limit of high Reynolds
number, that is $S_{p}(r)\sim(\frac{r}{L})^{\zeta_p}$ for $L >> r >>\eta_{k}$
, where $L$ is the scale of energy injection,
$\eta_{k}=(\frac{\nu^3}{\epsilon})^{1/4}$
is the dissipative scale, $\epsilon$ is the mean energy dissipation range
and $\nu$ is the kinematic viscosity.
Rare peaks in the random flows are signaled with a nonlinear form of $\zeta(p)$.
In other words the velocity increments are multifractal and $\zeta(p)$'s do not  
follow the celebrating K41 theory, $\zeta(p) = p/3$.
Recently \cite{lvov2,lvov3,eyink} it has been proposed that it would be
more natural to look at 
single time correlations among velocity increment fluctuations at different 
scales:
\bea 
{\cal F}_{n}(x|r_1,r_2,...,r_n)=\langle \delta_{r_1}u(x)
\delta_{r_2}u(x)...\delta_{r_n}u(x) \rangle \label{npoint}
\eea
where all the scales $r_i$ are lying in the inertial range,i.e. 
$\eta<<r_i<<L$
For simplicity we are confining the discussion to longitudinal
velocity increments. Fusion rules \cite{lvov2,lvov3,eyink} that describe
the asymptotic
properties of n-point correlation functions when some of the coordinates
tend toward one other are derived from two fundamental assumptions which
are of paramount importance for the description of non-perturbative
aspects of the analytic theory of stationary turbulence. The fusion rules
are tested experimentaly and a good agreement between experiment and
theory have been observed \cite{druhva}. 
If $p<n$ pairs of coordinates of velocity differences
merge, with the typical separations $r_{i} \sim r$ for $i \le p$
and the remaining separation at the order of $R$ such that $r<<R<<L$
, the fused multiscale correlation is defined as 
\bea
{\cal F}_{p+q}(r,R)&\equiv&\langle (u(x+r)-u(x))^{p}(u(x+R)-u(x))^{q} \rangle \nonumber \\
       &\equiv& \langle (\delta_{r}u(x))^{p} (\delta_{R}u(x))^{q} \rangle
\eea 
it has been deduced that 
\be 
{\cal F}_{p+q}(r,R) \sim S_{p}(r)
S_{p+q}(R)/S_{p}(R)\label{multiscale} 
\ee 
On the other hand recently
multiscale correlation functions in high Reynolds number experimental
turbulence, numerical simulations and synthetic signal was investigated by
Benzi and coworkers \cite{benzi} and it has been found that whenever the
simple
scaling ansatz based on the uncorrelated multiplicative processes
\cite{benzi}
is not prevented by symmetry arguments, the multiscale correlations are in
good agreement with the fusion rules prediction even if strong corrections
due to subleading terms are seen for small-scale seperation $r/R \sim
O(1)$ .  All the findings has led to the following conclusion that
multiscale correlation functions measured in turbulence are fully
consistent with a multiplicative almost uncorrelated random process for
evolution of velocity increments in scale .  Although successful
interpretation of the fusion rules can be realised by considering a
multiplicative random process for the evolution of velocity increments in
length scale, but it is at most a phenomenological modelling and it is not
based on a first principle calculations. Other front of experimental
investigations over the bahaviour of conditional probability densities of
velocity increments in scale have shown that Markovian nature of velocity
increments in terms of length scale and in the inertial range would
support the experimental data \cite{fp}. In fact the necessary condition
of
"Markovianity"  for velocity increments have been tested experimentaly
from which phenomenological senarios for modelling the intermittency have
been developed \cite{fp}. Afterwards the aforementioned ideas were
supported
by invoking to the theoretical ideas inspired by Polyakov \cite{pol} and
Yakhot \cite{ya} based on OPE and general invariances of the Navier-Stokes
equation \cite{ya}.
In this paper we show that the interpretation given by
Benzi and coworkers \cite{benzi} dictating an uncorrelated multiplicative
process for velocity increments in the inertial range is not just a mere
phenomenological interpretation and can be supported by Yakhot modelling. 
Even we will candidate the underlying dynamical process in scale which
incorporates the fusion rules of multiscale correlation functions. 
The calculations are
consistent with the picture of almost uncorrelated random multiplicative
process at least in the Fokker-Planck description of turbulence cascades.
Furthermore we are able to connect the fusion rules to the Markovian
nature by a simple operator formalism even by perserving all the terms in
the Kramers-Moyal's evolution operator of velocity increments.

Let us start with the Navier--Stokes equation:

\be
{\bf v}_t + ({\bf v} \cdot \nabla ) {\bf v} = \nu \nabla^2 {\bf v} - 
\frac {\nabla p}{\rho} + {\bf f}({\bf x},t), \hskip .5cm \nabla \cdot \bf{v}=0
\ee

for the Eulerian velocity $ {\bf v}({\bf x},t)$ and the pressure
$p$ with viscosity $\nu$, in N--dimensions.
The force $ {\bf f}({\bf x},t)$ is the external stirring force, which
injects energy into the system on a length scale $L$.
More specifically one can take, for instance a Gaussian distributed  
random force, which is identified by its two moments:

\be
\langle f_\mu ({\bf{ x}},t)  f_\nu ({\bf x^{'}},t^{'}) \rangle =  
k(0) \delta (t-t^{'}) k_{\mu \nu}({\bf {x}- { x^{'}}})
\ee
and $\langle f_\mu ({\bf{ x}},t) \rangle = 0 $,     
 where $\mu, \nu = x_1, x_2, \cdots ,x_N$. The correlation function 
$k_{\mu \nu}(r)$ is normalized to unity at the origin and decays 
rapidly enough where $r$ becomes larger or equal to integral scale $L$,
that is 
$ k_{\mu \nu} (r_{ij}) = k(0) [1- 
\frac {r_{ij} ^2}{2 L^2} \delta_{\mu,\nu} -
 \frac {({\bf r}_{ij})_\mu ( {\bf r}_{ij})_\nu}{L^2} ]$
with $k(0)$ and $ L \equiv 1$, where $ r_{ij} = | {\bf x_i} - {\bf x_j}|$.

Recently Yakhot \cite{ya} generelize the Polyakov's approach on 
the Burgers turbulence \cite{pol} for strong turbulence. He used the OPE
approach to closed the equation for the velocity increment PDF and show that
the PDF of longtudinal structure function $S_q = \langle (u(x+r) - u(x))^q \rangle =
\langle U^q \rangle$,
where $u(x)$ is the $x$-component of the three-dimensional velocity field
and $r$ is the displacement in the direction of the $x$-axis 
satisfy the follwing equation for homogeneous and isotropic turbulence
in the limit  $r \rightarrow 0$ as,

\be  
\frac{\partial}{\partial U} U \frac{\partial P}{ \partial r} -  B_0
\frac{\partial P}{ \partial r} 
= - \frac {A}{r} \frac{\partial} { \partial U} U P + \frac {u_{rms}}{L}
\frac{\partial^2}{ \partial U^2} U P 
\ee

where $A= \frac {3+B}{3}$, $B=-B_0 >0$ and  
for the Navier-Stokes turbulence it
has been shown that $B \sim 20$ can be derived by a self-consistent
calculations \cite{ya}.
Last term in the right hand side is responsible for the breakdown of 
Galilien invariance in the limited Polyakov sense, which means that the 
single-point $u_{rms}$ induced by random forcing enter the resulting 
expression for velocity increment PDF's.

Now one can show that the probablity density and as a result the conditional 
probablity density of velocity difference satisfy the Kramers--Moyal's  
evolution equation \cite{jhd}:
\be
-\frac{\partial{P}}{\partial{r}}=\sum_{n=1}^{\infty} {(-1)}^{n}\frac{\partial^{n}}{\partial{U}^{n}}(D^{(n)}(r,U)P)
\ee
where $D^{(n)}(r,U)=\frac{\alpha_{n}}{r} U^{n}+\beta_{n} U^{n-1}$.
We have found that the coefficients $\alpha_{n}$ and $\beta_{n}$ depend on $A$ and
$B$, $u_{rms}$ and inertial length scale $L$ which are given by 
$\alpha_n = (-1)^n \frac{A}{ (B+1) (B+2) (B+3) \cdots (B+n)} $ 
and $\beta_n = (-1)^n \frac {u_{rms}}{L} \frac{1}{(B+2)(B+3) \dots (B+n)}$ 
where $\beta_1 = 0$ by homogeneity \cite{jhd}.
The coefficients $D^{(n)}(r,U)$ are the small scale limit of the
conditional moments \cite{van}. They fully characterise the statsitics of 
eddy distribution in the inertial range which is defined as:
\begin{center} 
$D^{(n)}(U_2)= \lim_{r_1 \rightarrow r_2} \frac{1}{r_1-r_2} \int
(U_1-U_2)^{n}P(U_1,r_1|U_2,r_2)dU_1 $
\end{center}
It is noted that $P(U_1,r_1|U_2,r_2)$ statisfies the equation (7).
The Kramers-Moyal coefficients are the main observables of a Markov
process from which all the terms in the Kramers-Moyal operator will be
determined. It is a well known theorem (Pawula theorem) of Markov
processes that
whenever the fourth order Kramers-Moyal coefficient tends to zero all the
the other terms with higher order derivatives tend to zero \cite{van}.  
So the distinction between Markov processes which the Fokker-Planck
description when just the
first two terms in the evolution operator in scale are important and the 
Markov processes which all the terms should be preserved is related with
the coefficients too.   
Thanks to the detailed analysis carried over experimental data
\cite{fp}, the functional form of the first four Kramers-Moyal
coefficients has been obtained. It is been observed that the fourth
order conditional 
moment tends to zero from which by invoking to Pawula's
theorem the Fokker-Planck equation is canditated.  
In the mean time the present authors have shown that the Kramers-Moyal
coefficients which are derived from the Yakhot modelling are extremely
consistent with the experimental observations \cite{jhd}.
It is interesting that the functional forms of the different
coefficients $(D^{n}(U,r))$ are identically supporting the experimental
observations \cite{fp}.
The intermittency exponent of the structure functions can be derived from
equation (7), $\zeta(p)=\frac{Ap}{B+p}$.
It is easy to see that the ratio of different KM coeficients are
controlled by the $B$ parameter. As it is obvious when $B\rightarrow
\infty$ K41
scaling is recovered and $B \rightarrow 0$ produces the extreme case of
multiscaling
related to Burgers intermittency \cite{ya}. 
The Kramers-Moyal's description of PDF deformation in scale has been
supported by exact computation also for the compressible turbulence in
high Mach number limit \cite{tabar}. In that case the numerical values of
the $A$ and $B$ terms is determined without any needs to numerical
estimation.
Reminding the original idea for suggesting the Markovian property of 
energy cascade in scale we take a step further and calculate the 
more general objects of the cascade,i.e. the unfused multiscale
correlations. 
Assuming the Markovian nature of velocity increments in scale
and the proposed form of evolution operator $L_{KM}(U,r)$, one in
principle can calculate any correlation among velocity increments in
different scales:
\begin{center}
${\cal F}_{n}(x|r_{1},r_{2},...,r_{n})=\langle
U(r_{1})U(r_{2})...U(r_{n})\rangle
			 =\int dU(r_{1})...dU(r_{n}) U(r_{1})...U(r_{n})
P(U_{1},r_{1};U_{2},r_{2};...;U_{n},r_{n})$

\end{center}
The joint probability $P(U_1,r_1;U_2,r_2;...;U_n,r_n)$ can be calculated
by advantage of Markovian property in terms of conditional probabilities,
i.e.,
\bea
P&(&U_1,r_1;U_2,r_2;...;U_n,r_n)=P(U_1,r_1|U_2,r_2) \times \nonumber\\
\times
&P&(U_2,r_2|U_3,r_3)...P(U_{n-1},r_{n-1}|U_n,r_n)P(U_n,r_n)\label{rnpoint}
\eea
The conditional PDF of velocity increments can be written 
as a scalar--ordered operator as
$P(U_1,\lambda_1|U_2,\lambda_2)=
{\cal T}(e_{+}^{(\int_{\lambda_2}^{\lambda_1}
d \lambda L_{KM}(U_1,\lambda))})\delta(U_1 - U_2)$.
So in the calculation of n-point multiscale correlation a series of 
conditional operators would emerge in the integrand of (\ref{rnpoint}).
When some of the coordinates coalesce the conditional operator tends to 
a dirac delta function. The reduction of the conditional probability
between the coalescing coordinates simplify the calculations. The only
remaining conditional operator will be the probability of observing the
typical velocity $U_1$ increment between one subclass of fused points
conditioned on observing the typical velocity increment $U_2$ in the other 
subclass of fused points.
Explicitly we examine the behaviour of ${\cal F}_{p+q}(\lambda_1,\lambda_2)$
defined in \ref{npoint}, where $\lambda_1=\ln(L/r)$ and
$\lambda_2=\ln(L/R)$. 
\bea
{\cal F}_{p+q}(\lambda_1,\lambda_2)&=&\langle
U^{p}(\lambda_1)U^{q}(\lambda_2)\rangle=\int dU_1 dU_2 \delta(U_1-U_2)
\times \nonumber\\
\times &P&(U_2,\lambda_2)(e^{-(\lambda_1-\lambda_2)L_{KM}^{\dagger}(U_1)}
U_1^{p})U_2^{q}
\eea
We restrict the calculations to the GI invariant apprtoximation 
so that neglecting 
the $O(u_{rms} r/L)$ operators in $L_{KM}(U,\lambda)$. The    
crucial point in the above approximation is that in the GI regime 
the Kramers-Moyal's coefficients are scale independent so that all
the scale dependence of conditional probabilities would reveal as a
simple subtraction of the two logarithmic scales, i.e. $\lambda_1
-\lambda_2$ in the exponent. Because $L_{KM}^{\dagger}(U_1)
U_1^{p}=\zeta(p) U_1^{p}$, we will obtain the proposed form of the fusion
rules in (2) with $\zeta(p)=\frac{Ap}{(p+B)}$. Any other multiscale
correlation function is also tractable under the same approximations.\\
The Fusion rules first introduced \cite{lvov2,lvov3,eyink} through
invoking two
Kolmogorov type assumptions. The first one is the assumption of a
scale
invariant form for all the correlation functions in the inertial range.
The second is called the universality meaning that
when some arbitrary set of velocity differences in the correlation
functions are fixed in the scale $L$ , their precise choice will affect
the correlation functions just as an overall factor. In terms of
conditional averages the second proposition means that
\be
\langle U(r_1)^{p} | U(r_2)^{q} \rangle = S_{p}(r_1)\Phi _{p,q}(r_2) 
\ee
where it is assumed also that scale of $r_2$ is in the order of integral
scale while the $r_1$ is in the inertial range. 
The function $\Phi_{p,q}(r_2)$ isa homogenious function with a scaling
exponent $\zeta_n -\zeta_p$, and is associated with the remaining $n-p$
indices of ${\cal F}$. Mathematically the above
conditional correlation is easily verified 
\begin{center}
$\langle U(r_1)^{p} | U(r_2)^{q} \rangle = S_{p}(r_1) U_2^{p}/S_{p}(r_2)$
\end{center}
In Yakhot modelling scaling hypothesis has been invoked from the very
beginning of the theory when the relevant OPE terms has been chosen for 
closing the equation governing over the generating function of
longitudinal velocity increments. But we are showing that 
at least in the framework of Yakhot modelling 
the universality 
proposition is the {\it result} of the Markovianity of the evolution of
velocity increments in scale. On the other hand the necessary
proof of Markovian property 
has been verified through the special scalar-ordered form of the
conditional probabilities. This itself has arisen from the general
invariances and scaling constraint of the Navier-Stokes Equation. 
So the universality condition in the language of multiscale correlation
functions has in it's heart the very robust scaling invariance under the 
infinite parameter scaling group \cite{fri}.
We should emphasise that non-universal effects of the large scale motions 
can also manifest through the scale dependent terms in the Kramers-Moyal
operator. Still the general form of Universality assumption would be the 
leading behaviour while the $O(u_{rms}r/L)$ will be the  
sub-leading corrections inducing the large scale effects \cite{dr}.\\
Within the experimentally verified approximation that neglects the
third
and
higher order KM coefficients \cite{jhd,fp}, one can write the equivalent
diffusion
process in scale which dynamically gives the relation between velocity 
increments at two different scales. In fact approximating the KM equation
with a Fokker-Planck evolution kernel can be interpreted as if velocity
increment $U$ is evoluted in "scale" $\lambda$ (logarithmic length scale)
by the langevin equation \cite{van},
$\frac{\partial{U}}{\partial{\lambda}}=\tilde{D}^{(1)}(U,\lambda)+\sqrt{\tilde{D}^{(2)}(U,\lambda)}\eta(\lambda)$
, where $\eta(\lambda)$ is a white noise and the diffusion term acts as a 
multiplicative noise. 
Giving address to the Ito's prescription \cite{van} the multipoint
correlation function can be written in the form of a
path integral as
\bea
{\cal F}(\lambda_1,\lambda_2) = \int {\cal D}U U^{p}(\lambda_1&)&
U^{q}(\lambda_2)
e^{\int_{\lambda_1}^{\lambda_2}(\frac{\frac{\partial
U}{\partial \lambda}-D^{1}(U,\lambda)}{\sqrt{D^{2}(U,\lambda)}})d\lambda}
\times \nonumber\\
\times &P&(U_2,\lambda_2)
\eea
By a simple application of Bayesian rule probability density in the outer 
scale $\lambda_2$ can also be written as a path integral entering the
information of the integral scale PDF which in a good approximation can be
regarded as a Gaussian distribution \cite{ya}. Building up all the terms 
in a descriptive way the joint probability
$P(U_1,\lambda_1;U_2,\lambda_2)$ is represented as a path integral over
all possible {\it paths} between $U(\lambda_1)$ and $U(\lambda_2)$
transfering in an intermittent way all the information of the integral
scale into the calculation.  
Without further attempt for calculating the multiscale correlation
by path integral representation, 
we turn our attention to the Langevin dynamics instead.
The resulting
process is the well known Kubo \cite{van} oscillator multiplicative
process.
By using the Ito \cite{van} prescription one can deduce that 
\be 
\delta_{\lambda_1}U(x)= {\cal W}(\lambda_1,\lambda_2)
\delta_{\lambda_2}U(x)\label{mult}
\ee
The multiplier ${\cal W}(\lambda_1,\lambda_2)$ can be easily derived
in terms of
$\alpha_1$ and $\alpha_2$ and Wiener process at two logarithmic scales
as: 
\begin{center}
${\cal
W}(\lambda_1,\lambda_2)=e^{(-\alpha_{1}(\lambda_1-\lambda_2)
+\sqrt{\alpha_{2}}[W(\lambda_1)-W(\lambda_2)])}$.
\end{center}
Equation (\ref{mult}) encodes a simple cascade process.
Cascade processes are simple and well known useful tools to describe the
leading phenomenology of the intermittent energy transfer in the inertial
range. Both anomalous scaling exponents and viscous effects \cite{fri} can
be 
Under the framework of the derived KM equation we have shown that the
equivalent cascade model from Navier-Stokes equation is derived by cutting
the KM evolution operator after third term.
In fact we have given a recipe for finding the equivalent approximate
cascade model from Navier-Stokes equation.
Structure functions are described in terms of multiplier
$W(\lambda_1,\lambda_2)$ through $S_{p}(r)= C_{p} \langle [{\cal W}(r/L)]^{p} \rangle$
where from
the Langevin equation pure power law arise in the high Reynolds regime
$\langle [{\cal W}(r/L)]^{p} \rangle \sim(r/L)^{\zeta(p)}$. In this
approximation the scaling exponents
would be $\zeta(p)=-p \alpha_1 + p(p-1)\alpha_2/2$. From the direct
calculation of Langevin equation one can easily find the behavior of the
multiscale correlation function ${\cal F}_{p+q}(r,R)$. In the same
framework, it
is staightforward to show that 
\bea
{\cal F}_{p+q}(r,R)\sim\langle [{\cal W}(r,R)]^{p} [{\cal W}(R,L)]^{q}
\rangle \nonumber\\
\sim \langle [{\cal W}(\frac{r}{R})]^{p}\rangle \langle[{\cal W}(\frac{R}{L})]^{q}\rangle
\sim S_{p}(r) S_{p+q}(R)/S_{p}(R)
\eea
The independence of multipliers in two
diferent scales is allways assumed for the underlying cascade process
otherwise the following relation would not be held. Our modelling
equivalently encodes the following requirement by the obvious independency
of increments in a Wiener process. 
Recently Benzi and coworkers \cite{benzi} analysed multiscale correlation
functions from high Reynolds experiments and synthetic signals. They 
have elegantly seek whether the fusion rules (\ref{multiscale}) are
compatible with the random cascade phenomenology. Their main result is
that all multiscale correlation functions are well reproduced in their
leading term $\frac{r}{R} \rightarrow 0$ by a simple uncorrelated random
cascade.
In Yakhot modelling of the dynamics of the longitudinal velocity
increments in scale all the above results are recovered in the framework
of Fokker-Planck approximation. 
It is also interesting also to seek the limiting behaviour of the
Multiscaling correlation function for Burgers turbulence which is
tractable by taking the limit of $B\rightarrow 0$ in our formulation.
Equation (3) shows that multiscaling correlation function will be
independent of the outer scale $R$ which is consistent with our knowledge
about Burgers turbulence \cite{pfrisch}. 
We think that preserving all the terms in KM equation would give the full
information of cascade in length scale and this would give an answer to
the question that wether there are other subleading processes acting for
energy transfer from large to small scales. Preserving the GI breaking
terms in the corresponding stochastic processes would also give the
important unanswered question regarding the effect of uneven PDF's of
velocity increments on the cascade models. 
\vskip +0.1cm

We thank Uriel Frisch, Itammar Procaccia and Victor Yakhot for useful
remarks and comments. M.R. Rahimi Tabar wish to acknowledge 
Observatoire de la C$\hat o$te d'Azur where some part of this work is
done and U. Frisch for the kind hospitality.  

\vskip -.5cm


\begin{thebibliography}{99} 
\bibitem{fri} U. Frisch " Turbulence" Cambridge University Press (1995)
\bibitem{lvov2} V.S.L'vov and I. Procaccia, Phys.Rev.Lett. {\bf 76}, 2898
(1996).
\bibitem{lvov3} V.S.L'vov and I. Procaccia, Phys.Rev. E {\bf 54}, 6268
(1996).
\bibitem{eyink} G. Eyink, Phys.Lett. A {\bf 172}, 355 (1993);
G. Eyink, Phys.Rev. A {\bf 48}, 1823 (1993).
\bibitem{druhva} A.L. Fairhall, B. Druhva, V.S. L'vov, I. Procaccia, and
K.S. Sreenivasan( to be published).
\bibitem{benzi} R. Benzi, L. Biferale, and F. Toschi, Phys.Rev.Lett. {\bf
80}, 3244 (1998).
\bibitem{fp} R. Friedrich, J. Peinke, Phys. Rev. Lett. {\bf 78},863 (1997)
;B.Chabaud, A. Naert, J. Peinke,
F. Chilla, B. Castaing and B. Hebral, Phys. Rev. Lett. {\bf 73}, 3227
(1994).
\bibitem{pol} A. Polyakov, Phys. Rev. E {\bf 52}, 6183 (1995)
\bibitem{ya} V. Yakhot, Phys. Rev. E, {\bf57}, 1737 (1998)
\bibitem{van} N. G. van Kampen,"Stochastic Prosses in Physics and
Chemistry(Elsevier, Amesterdam 1990); H. Risken, " The Fokker--Planch Equation"
(Springer-Verlag, 1984)
\bibitem{jhd} Jahanshah Davoudi and M. R. Rahimi Tabar," Phys.Rev.Lett. 
{\bf 82},
1680 (1999). 
\bibitem{tabar} M.R.Rahimi Tabar, Jahanshah Davoudi, A. Rastegar, IPM
Preprint.
\bibitem{pfrisch} U. Frisch, private communication.
\end{thebibliography}
\end{document}